\documentclass[reprint, superscriptaddress,amsmath,amssymb, aps, pre]{revtex4-1}

\usepackage{graphicx}
\usepackage{bm}
\usepackage[usenames]{color}
\usepackage{epstopdf}
\usepackage{wasysym}
\usepackage{ulem}
\usepackage{geometry}
\usepackage{xcolor}
\begin{document}

\title{How capillarity affects the propagation of elastic waves in soft gels}
\author{P.~Chantelot}
\thanks{These two authors contributed equally}
\author{L.~Domino}
\thanks{These two authors contributed equally}
\author{A.~Eddi}

\affiliation{PMMH, CNRS, ESPCI Paris, PSL University, Sorbonne Universit\'e, Universit\'e de Paris, Paris, France
}

\begin{abstract}
Elastic waves propagating at the interface of soft solids can be altered by the presence of external forces such as capillarity or gravity.
We measure the dispersion relation of waves at the free surface of agarose gels with great accuracy, revealing the existence of multiple modes as well as an apparent dispersion.
We disentangle the role of capillarity and elasticity by considering the 3D nature of mechanical waves, achieving quantitative agreement between theoretical predictions and experiments.
 Notably, our results show that capillarity plays an important role for wavenumbers smaller than expected from balancing elastic and capillary forces.
   We further confirm the efficiency of our approach by including the effect of gravity in our predictions and quantitatively comparing it to experiments.
\end{abstract}

\maketitle

\section{Introduction} Mechanical waves propagating in biological tissues have been at the center of attention since the development of ultrasonic imaging more than 50 years ago \cite{wild1952}.
Using soft materials to mimic the physics of wave propagation inside the body has enabled to develop technological innovations, such as elastography, allowing for a direct measurement of the bulk elastic properties \cite{sandrin1999,bercoff2004}.
Soft solids have also been used as a model for fracture dynamics \cite{bonn1998,livne2005} and, in particular, for the role of friction and fault structure on rupture dynamics during earthquakes \cite{latour2011, latour2013}.
Wave propagation at interfaces raises the question of additional forces competing with elasticity. Indeed, solid interfaces possess a surface tension $\gamma$ that dominates bulk elasticity at small scale, below the elastocapillary length $\ell_{ec} = \gamma/\mu$ where $\mu$ is the solid shear modulus \cite{snoeijer2016,style2017,bico2018}. Depositing liquid drops on soft substrates allows to probe the competition between elasticity and capillarity, as the wetting ridge induced by the contact line sets the drop's statics and dynamics \cite{andreotti2019}.
For very soft solids, $\ell_{ec}$ can be as large as 1 mm. Capillary phenomena then become macroscopically visible at free surfaces: edges are rounded \cite{hui2002} and cylinders develop undulations reminiscent of the classical Plateau-Rayleigh instability for liquids \cite{mora2010}.
Waves existing at the interface of soft materials have been only partially described so far. The existence of two regimes, dominated by elasticity or capillarity, theoretically predicted \cite{harden1991} and initially probed experimentally in the late \mbox{-90s} \cite{monroy1998} has been at center of discussion \cite{onodera1998,ahn2001}.
Recent work focussed on the transition between the two regimes, yet with limited experimental resolution \cite{shao2018}. In this article, we propose to combine accurate wavefield measurements and a theoretical analysis in order to discriminate the influence of capillarity on the propagation of mechanical waves at the free surface of soft gels.

\begin{figure*}[t]
\centering
 \includegraphics[width=1\textwidth]{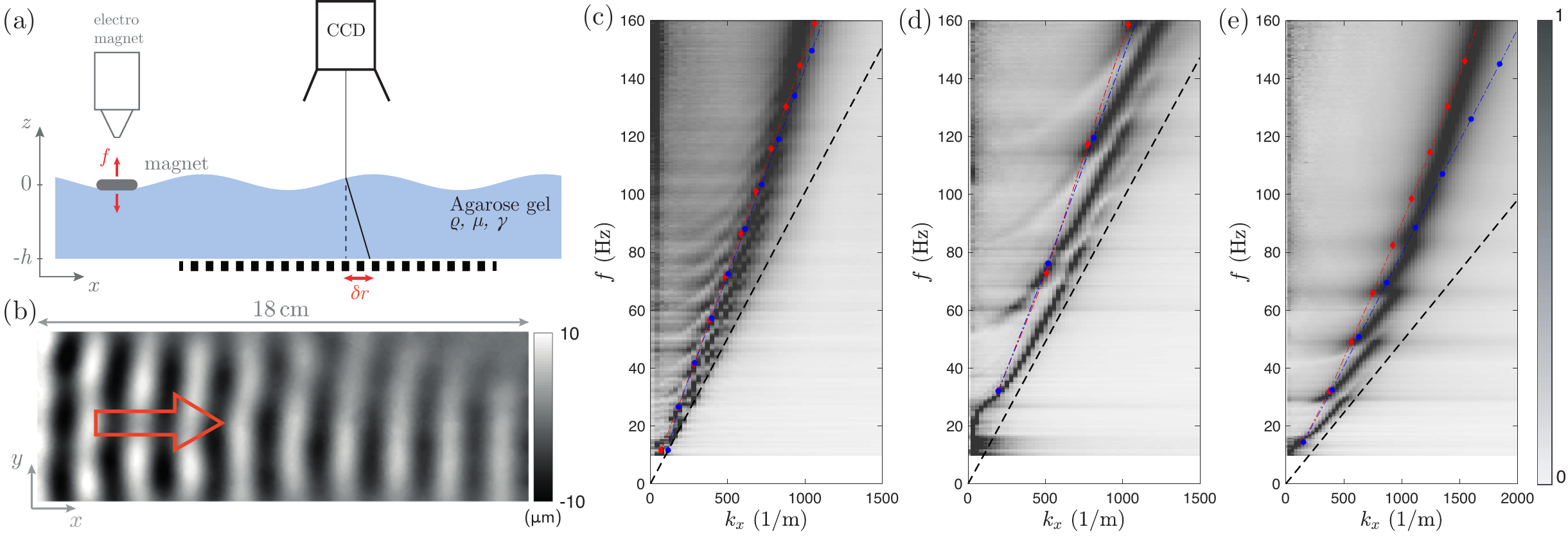}
   \caption{
   (a) Sketch of the experimental setup using SSI \cite{wildeman2018}.
   (b) Height field obtained for $f=40$ Hz in a gel with $\mu=95$ Pa and $h=1.1\pm0.1$ cm. The source is on the left and the red arrow shows the direction of propagation.
   (c-d-e) Dispersion relations measured at the gel interface, the wavenumber is measured along the $x$ axis. The black dashed lines show the dispersion relation of shear waves: $\omega=k\sqrt{\mu/\rho}$. The red and blue symbols represent the maximum of the normalized out-of-plane displacement along each mode predicted by equation \ref{disprel} with or without taking into account capillarity.
(c) $\mu = 380$ Pa, $h= 3.4 \pm 0.4$ cm.
(d) $\mu = 380$ Pa, $h= 1.1 \pm 0.1$ cm
(e) $\mu = 95$ Pa, $h = 1.1 \pm 0.1$ cm.
}
   \label{fig:SchemaHydrogelPiV}
\end{figure*}

\section{Experimental setup} We make agarose gels by heating a solution of water and agarose (Sigma A4550-500G) at $95^\circ$C. The solution is poured in a rectangular container ($8.5\times26$ cm) and left to cool at room temperature for 2 hours.
We determined the rheology of the hydrogels for concentrations of 2 g/L and 3 g/L which gives shear moduli, $\mu$, of respectively 95 Pa and 380 Pa (see appendix \ref{apprheol}). We used the samples within one hour after reticulation, so that evaporation does not affect their mechanical properties (see appendix \ref{appevap}).
We generate plane waves at the air/gel interface by locally imposing a vertical sinusoidal motion with frequency $f$ at the free surface of the sample. To do so, we deposit a rectangular source with dimensions $80\times8\times2$~mm, which size does not influence the results (see appendix \ref{appfinitesize}), on the surface of the gel and actuate it with an electromagnet, or alternatively we use a vibration exciter (figure \ref{fig:SchemaHydrogelPiV}a).
Several techniques have been used in the literature to measure surface waves, including quasielastic surface light scattering \cite{monroy1998}, specular reflection spectroscopy \cite{tay2008, pottier2011} or oscillatory response of a magnetic exciter \cite{rolley2019}.
The proposed methods are particularly adapted for short wavelengths, whereas here we want to measure extended wavefields with centimetric to millimetric wavelengths.
We thus choose to measure the out-of-plane displacement field at the interface using a Synthetic Schlieren Imaging (SSI) technique, based on the apparent displacement of a pattern caused by refraction at the surface \cite{wildeman2018}.
We record from the top at a frame rate of 350 Hz for $f$ ranging from 10 Hz to 160 Hz. We use sweeps at a rate of 1.6 Hz/s, small enough to consider the excitation as monochromatic when analysing small signal windows.

\section{Dispersion relation} We show in figure \ref{fig:SchemaHydrogelPiV}b a typical height field obtained at $f=40$~Hz in a gel with $\mu=95$~Pa and depth $h=1.1\pm 0.1$ cm.
We extract the wavefield at any frequency by Fourier filtering a signal window around the corresponding $f$. We then use spatial 2D Fourier transforms to extract the spectra along the propagation direction that we normalize by their maximum amplitude.
By stacking the spectra obtained at each $f$, we can construct a dispersion relation map, which we show in figure \ref{fig:SchemaHydrogelPiV}c for a gel with $\mu=380$ Pa and thickness $h=3.4\pm 0.4$ cm.
It shows the coexistence of two distinct behaviors. (i) For $f<120$ Hz, we observe multiple branches, which start at increasing cutoff frequencies.
(ii) At higher frequencies, the branches merge, and a single line is observed. We interpret the presence of several cutoff frequencies (at $k_x=0$) as a signature of the finite thickness: in a confined sample the vertical component of the wave vector can only take discrete values.
We investigate this effect by decreasing $h$ to $1.1\pm 0.1$ cm while keeping $\mu$ constant (fig. \ref{fig:SchemaHydrogelPiV}d). We observe the strong effect of the depth: there still are several branches but with markedly different cutoff frequencies. The fundamental mode appears at a higher frequency, the following branches start existing at larger $f$ and are further apart.
Then, we probe the effect of the gel's elastic properties by decreasing the agarose concentration to obtain a gel with $\mu=95$ Pa and $h=1.1$~cm (fig. \ref{fig:SchemaHydrogelPiV}e). The cutoff frequencies are now lower.
We note that the local slope of each branch is significantly smaller than that of the stiffer gels (fig. \ref{fig:SchemaHydrogelPiV}c and \ref{fig:SchemaHydrogelPiV}d).
Plotting on figures \ref{fig:SchemaHydrogelPiV}c-e the dispersion relation of shear waves $\omega = k\sqrt{\mu/\rho}$ (black dashed-lines) \cite{landau1986} suggests that this local slope is controlled by the speed of elastic shear waves.
Conversely, the slope of the single line observed at high frequency is larger than that of shear waves. The dispersion relation can be regarded as an apparent dispersion curve whose group velocity progressively increases.
The latter effect, as well as the increase of the local slope of the branches at high $k$ in the softer gel (fig. \ref{fig:SchemaHydrogelPiV}e), both hint at the presence of capillarity that could stiffen the interface at large $k$.

\section{In-depth displacements} 
\begin{figure}[t!]
\centering
\includegraphics[width=0.5\textwidth]{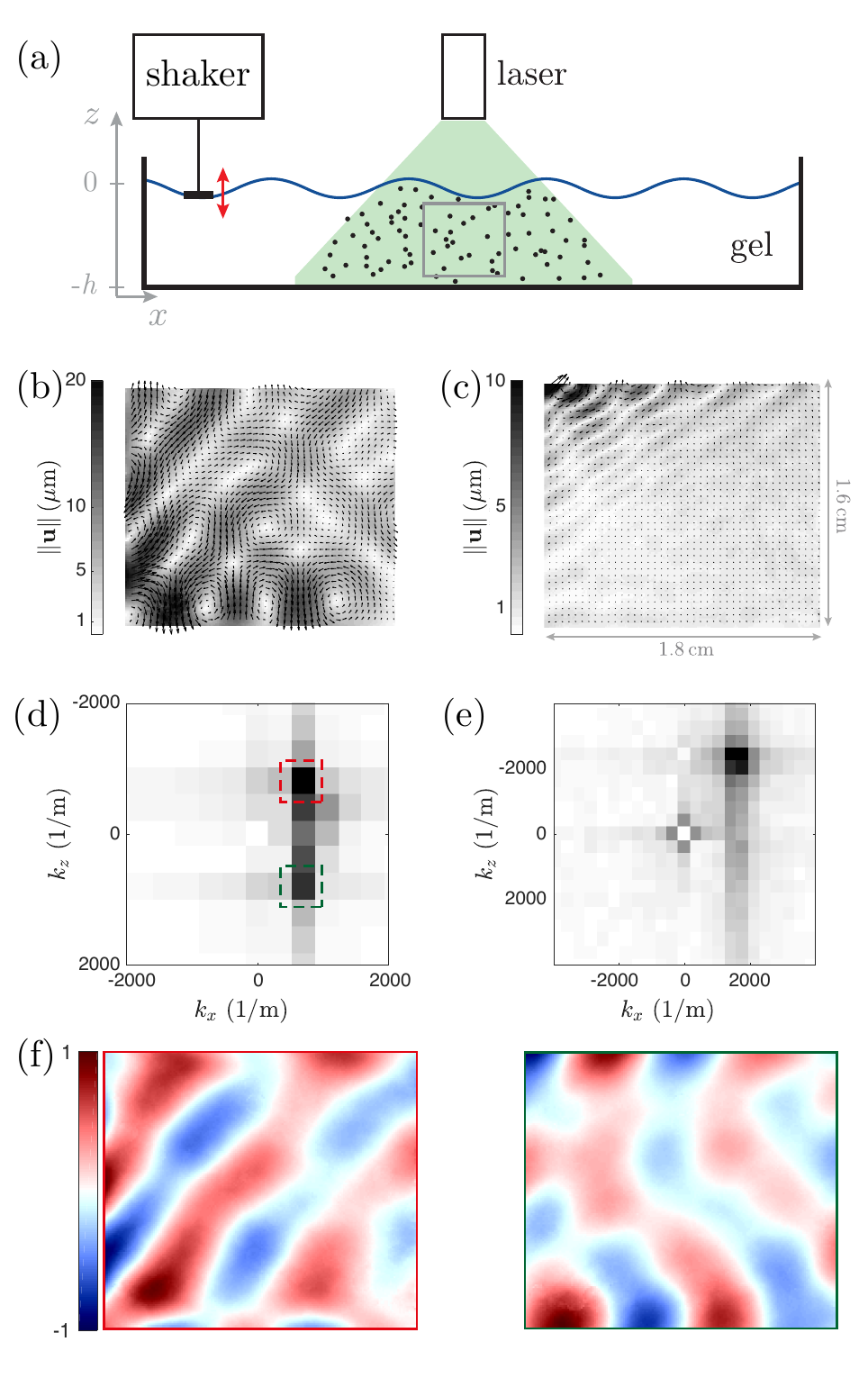}
\caption{\label{figPIV}(a) Sketch of the DIC experimental setup used to measure in-depth displacement fields. (b-c) Displacement field inside a gel with $\mu = 95$ Pa and $h = 2.3 \pm 0.3$ cm for (b) $f=40$ Hz and (c) $f=120$ Hz. (d-e) Spatial spectra corresponding to the fields in (b-c). The two peaks in (d) correspond to the presence of an incident wave and its reflection at the bottom of the tank. (f) Wave fields obtained by taking the inverse Fourier transform of each peak in (b)}
\end{figure}

Surface measurements suggest that the finite thickness selects the modes at low $f$. We confirm this hypothesis by measuring in-depth displacement fields. We seed the gel with micro-particles (diameter $\sim10$ $\mu$m, density 1100 kg/m$^3$) and illuminate the $xz$ plane with a laser sheet (fig. \ref{figPIV}a). We use a low micro-particle concentration, $\chi = 0.14 \%$, so that the inclusions do not modify the gel elastic properties (see appendix \ref{appinclus}).
We measure the local displacement field at 250 fps, using a standard Digital Image Correlation (DIC) algorithm \cite{blaber2015}, in a window with dimensions $1.8\times1.6$ cm approximately $2$ cm away from the source.
Figures \ref{figPIV}b-c present a quiver plot of the displacement vector superimposed over a map of its magnitude for a gel with $\mu=95~$Pa and $h=1.9 \pm 0.1$ cm excited at $f=40~$Hz (fig. \ref{figPIV}b) and $f=120~$Hz (fig. \ref{figPIV}c).
The displacement amplitude is on the order of $10~\mu$m, and both vertical and horizontal components are present.
For $f=40$ Hz, we observe displacements in the entire sample, without a significant decay in the vertical direction while at $f = 120$ Hz the amplitude seems to decrease faster in the vertical than in the horizontal direction.
We extract the spatial spectra corresponding to these displacements fields (fig. \ref{figPIV}d-e). For both frequencies, the wave vectors have a non-zero component on the vertical axis: the previous surface measurements correspond to their horizontal projection. The norm of the wave vector is $||\mathbf{k}|| = 981$ 1/m for $f= 40$ Hz and $||\mathbf{k}|| = 2768$ 1/m for $f= 120$ Hz, two values compatible with the propagation of shear waves in the bulk ($\omega= k\sqrt{\mu/\rho}$).
Yet, the two spectra are markedly different. For $f=40$ Hz, we observe two peaks that correspond to the presence of an incident ($k_z<0$) and reflected ($k_z>0$) wave created by the reflection at the bottom of the tank. We evidence this result by plotting in figure (fig. \ref{figPIV}f) the wave fields obtained by taking the inverse Fourier transform of each peak.
At higher frequency ($f = 120$ Hz), the spatial spectrum shows only one peak (fig. \ref{figPIV}d).
The wave travelling downwards is damped before it reaches $z=-h$, so that propagation occurs mostly at the surface.
These experiments 
confirm that the multiple modes observed at low frequency result from the vertical confinement and they suggest that, at high frequency, dissipation prevents the incident wave to propagate all the way down to the bottom of the sample.

\begin{figure*}[t]
\centering
\includegraphics[width=1\textwidth]{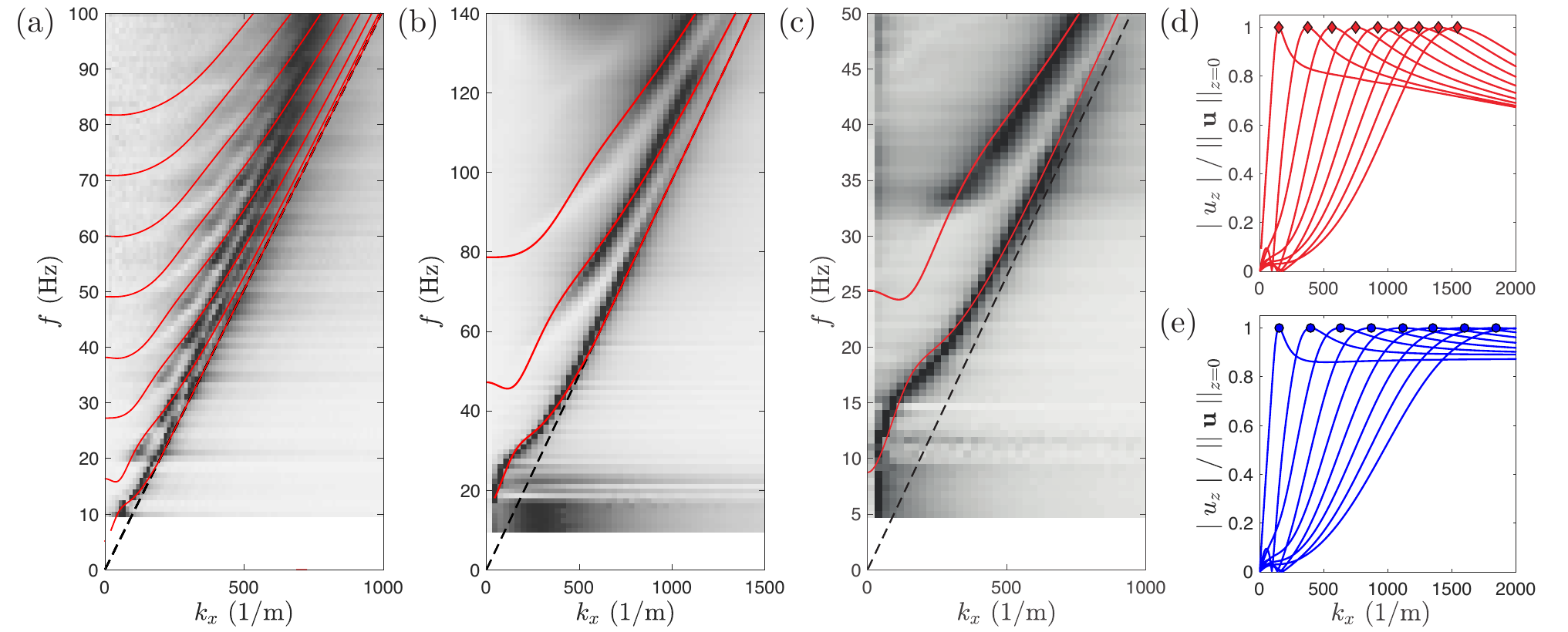}
\caption{\label{fig3} (a-c) Overlay of the dispersion maps measured on samples with different $\mu$ and $h$ and the dispersion curves obtained by computing the zeros of equation \ref{disprel} ($\mu_{th}$ being the only adjustable parameter). The dashed lines show the dispersion relation of shear waves: $\omega=k\sqrt{\mu/\rho}$. (a) $\mu = 380$ Pa, $h = 2.90 \pm 0.05$ cm, $\mu_{th} = 400$ Pa (b) $\mu = 380$ Pa, $h = 0.98\pm 0.05$ cm, $\mu_{th} = 380$ Pa and (c) $\mu = 95$ Pa, $h=0.99 \pm 0.05$ mm, $\mu_{th}=110$ Pa. (d-e) Normalized vertical displacement $| u_z |/||\mathbf{u}||$ as a function of $k$ for $\mu_{th}=120$ Pa and $h_{th}=1.3$ cm with (d, red lines) and without (e, blue lines) taking into account capillarity. }
\end{figure*}

\section{Modeling}  We now model wave propagation in soft materials.
To account for our experimental observations, we address the case of vertically confined samples.
We extend previous analysis \cite{onodera1998}, that treated the case of a semi-infinite solid subject to elastic and capillary forces, to a finite thickness sample and add the contribution of gravity.
We consider plane waves propagating along the $x$ direction in an infinite 2D plate of thickness $h$, density $\rho$ with elastic properties characterized by the Lam\'e coefficients $\lambda$ and $\mu$. We separate the displacement field $\bm{u}$ in a longitudinal curl free contribution $\bm{u}_l$ and in a transverse divergence free contribution $\bm{u}_t$. The longitudinal part can be described by a scalar potential $\Phi$ and the transverse part by a vector potential $\bm{H}$.
\begin{equation*}
\bm{u} = \bm{u}_l + \bm{u}_t = \nabla \Phi + \nabla \times \bm{H}.
\end{equation*}
Both $\Phi$ and $H_y$ verify a wave equation \cite{landau1986}:
\begin{equation*}
\nabla^2\Phi -\frac{1}{c_l^2}\frac{\partial^2 \phi}{\partial t^2} = 0,\, \nabla^2 H_y -\frac{1}{c_t^2}\frac{\partial^2 H_y}{\partial t^2} =0
\end{equation*}
where $c_t = \sqrt{\frac{\mu}{\rho}}$ and $c_l=\sqrt{\frac{\lambda+2\mu}{\rho}}$ are, respectively, the shear and longitudinal wave speeds.
We seek solutions of the form $\Phi = f(z)e^{i(kx-\omega t)}$ and $H_y = ih(z)e^{i(kx-\omega t)}$ where we use $k$ as $k_x$ and impose the following boundary conditions.
(i) At the bottom of the sample, the gel is bounded to the container, so that:
\begin{equation*}
u_x(z=-h)=u_z(z=-h)=0.
\end{equation*}
(ii) At the free surface, assuming small deformations to linearize the boundary conditions at $z=0$ and taking advantage of the incompressibility of the hydrogels ($c_l\rightarrow\infty$) that allows to absorb bulk gravity into the hydrostatic pressure, we impose:
\begin{equation*}
 \sigma_{xz}(z=0) = 0,\, \sigma_{zz}(z=0) = \gamma \frac{\partial^2 u_z}{\partial x^2}-\rho g u_z.
\end{equation*}
Only the boundary condition at the interface sets this problem apart from the purely elastic one: capillarity  and gravity are taken into account by respectively relating the Laplace and hydrostatic pressure to the normal stress.
Using the four boundary conditions, and substituting $\Phi$ and $H_y$, we obtain the dispersion relation for the gravito-elasto-capillary waves (see appendix \ref{appmodeling}). This relation can be written in dimensionless form by introducing the variables $\tilde{k}=kh$ and $\tilde{\omega} =\omega h /c_t$:
\begin{multline}
\tilde{k}^{2}\sinh\tilde{\alpha}\sin\tilde{\beta}\left((\tilde{k}^{2}-\tilde{\beta}^{2})^2 + 4\tilde{\alpha}^{2}\tilde{\beta}^{2}\right) \\
- \tilde{\alpha}\tilde{\beta}\cosh\tilde{\alpha}\cos\tilde{\beta} \left(4\tilde{k}^{4}+(\tilde{k}^{2}-\tilde{\beta}^{2})^2\right) \\
+ 4\tilde{\alpha}\tilde{\beta}\tilde{k}^{2}(\tilde{k}^{2}-\tilde{\beta}^{2}) - \left( \Gamma + \frac{G}{\tilde{k}^{2}}\right)\tilde{\alpha}\tilde{k}^{2}\left(\tilde{k}^{2}+\tilde{\beta}^{2}\right) \\
\left(\tilde{k}^{2}\cosh\tilde{\alpha}\sin\tilde{\beta} +\tilde{\alpha}\tilde{\beta}\sinh\tilde{\alpha}\cos\tilde{\beta}\right) =0
\label{disprel}
\end{multline}
where $\tilde{\alpha}^{2} = - \tilde{k}^{2}$ and $\tilde{\beta}^{2} = \tilde{\omega}^{2} - \tilde{k}^{2}$.
We identify two dimensionless parameters $\Gamma = \gamma/\mu h$ and $G = \rho g h/\mu$ that compare the elastocapillary length $\ell_{ec}=\gamma/\mu$ and the elastogravity length $\ell_{eg}=\mu/\rho g$ to the thickness $h$.
Using a secant method algorithm, we determine the zeros of equation \ref{disprel} (assuming that the surface tension of the gels is similar to that of water, \textit{i.e.} $\gamma=70$ mN/m). In figures\ref{fig3}a-c, we overlay the obtained curves (red lines) on experimental maps where the thickness of the sample was precisely controlled (so that $\mu$ is the only adjustable parameter in equation \ref{disprel}).
The model is in good agreement with the measured data: it captures the existence of multiple branches, their cutoff frequencies and local slope when varying both $\mu$ and $h$.
The values $\mu_{th}$ used to fit the predicted relations to the experimental data are always larger than the expected $\mu$. We qualitatively ascribe this discrepancy to the evaporation of the hot agarose solution during preparation \cite{evap}.
Yet, we do not explain the apparent dispersion: the signal is localized on a finite part of the predicted branches.
We gain more insight by deriving the displacement field associated to each mode: for any ($\omega$, $k$) verifying equation \ref{disprel}, we can compute the displacement field at the interface up to a multiplicative constant (see appendix \ref{appmodeling}). We plot in figure \ref{fig3}d the norm of the vertical displacement normalized by the magnitude of the displacement vector at the interface, $| u_z |/||\mathbf{u}||_{z=0}$, as a function of $k$ (red lines) for $\mu_{th}=120$ Pa and $h_{th}=1.3$ cm.
The normal displacement at $z=0$ varies in a similar fashion for each mode: it increases sharply until it reaches a maximum  for $k=k_m$ (red diamonds) and then decreases at a slower rate.
As SSI only detects out-of-plane motion, we expect to measure waves only when $k>k_m$ and that the signal intensity decays along each mode as $k$ increases.
We report with red symbols in figure \ref{fig:SchemaHydrogelPiV}c-e the couples ($\omega_m$, $k_m$) obtained from the model for each sample. Our prediction now captures the apparent dispersion, the red diamonds act as lower bounds for the presence of signal for each mode.
The blurring of the modes into a single line can be qualitatively explained by the significant effect of dissipation at high frequency, an effect that would deserve a separate study.

\section{Elasto-capillary effect}
Although the shape of the apparent dispersion suggests that it is caused by surface tension, balancing the capillary induced stress, of order $\gamma k$, with the elastic stress predicts that capillarity dominates when $k > 2\pi/\ell_{ec} = 8.5\cdot10^3 ~\textrm{m}^{-1}$ (for $\mu=95$~Pa), much larger than the wavenumbers probed experimentally.
We report in figure \ref{fig3}e the normalized vertical displacement at the interface for the same parameters as in figure \ref{fig3}d without taking into account surface tension to evidence its role.
The variations of the out-of-plane displacement are different when $k > k_m$, where we now observe a plateau.
The nature of the displacement fields is modified, reducing the relative weight of the out-of-plane contribution. Physically, an extra energetical contribution due to capillarity tends to favour in-plane displacements even for $k < 2\pi/\ell_{ec}$.
We also notice that the values of ($\omega_m$, $k_m$) are shifted so that we no longer recover the apparent dispersion in figures \ref{fig:SchemaHydrogelPiV}c-e (blue circles): they align on a line with slope $\sqrt{2}c_t$, corresponding to Lam\'e modes (see appendix \ref{appLame}).
This shows that the apparent dispersion is caused by capillarity for wavenumbers lower than $2\pi/\ell_{ec}$.
Since $\Gamma$ ranges from 0.001 to 0.08 in the experiments of figure \ref{fig:SchemaHydrogelPiV} and figure \ref{fig3}, the shape of the predicted modes is hardly modified by capillarity.
To probe the effect of capillarity on the dispersion curves, we investigate wave propagation in a very shallow sample ($\Gamma \sim 1/h$). We report in figure \ref{fig4}a the dispersion relation of a gel with $\mu=95$ Pa and $h=0.23\pm 0.05$ cm for which $\Gamma=0.4$. The red lines represent the prediction of equation \ref{disprel} with (solid line) and without (dashed line) capillary effects.
The prediction lies closer to the experimental result when including capillarity, which confirms its direct influence.
It is worth noticing that the two effects discusssed above are specific to finite thickness configurations and are markedly different from the elastic to capillary transition discussed in \cite{monroy1998,shao2018}.

\section{Elasto-gravity effect}
\begin{figure}[t]
\centering
\includegraphics[width=0.5\textwidth]{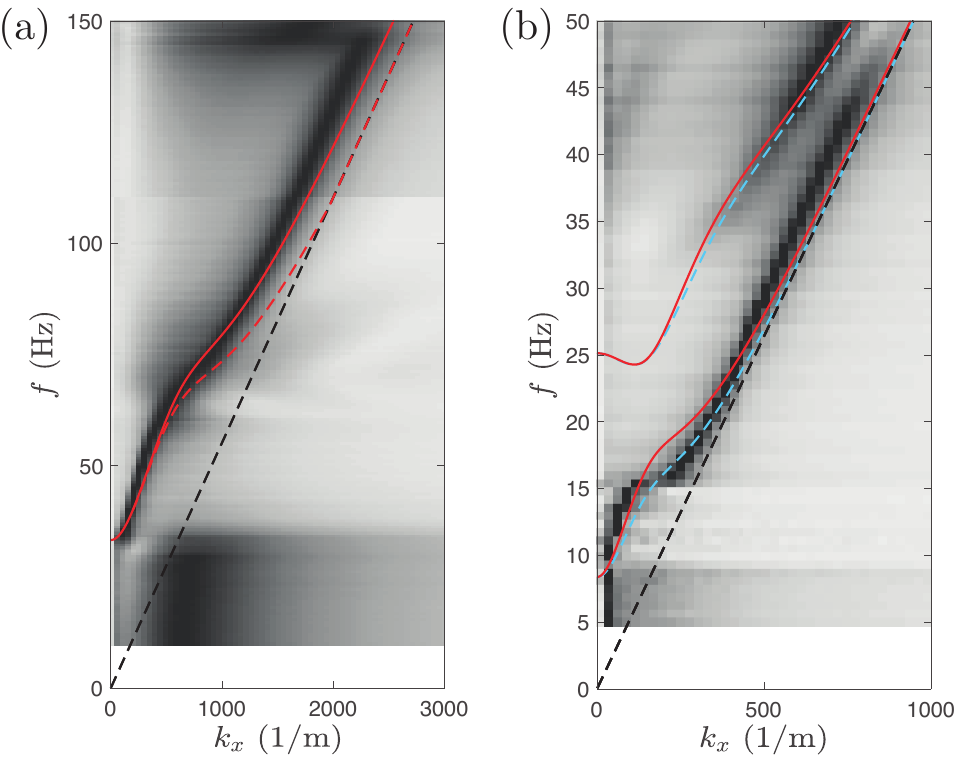}
\caption{\label{fig4}(a) Dispersion map obtained for $\mu = 95$ Pa and $h=0.23\pm 0.05$ cm. The red solid (resp. dashed) line represent the prediction of equation \ref{disprel} ($\mu_{th} = 120$ Pa, $h_{th}= 0.26$ cm) with (resp. without) including capillarity. (b) Dispersion map for $\mu=95$ Pa and $h=0.99\pm0.05$ cm when the interface points downwards. The red solid (resp. cyan dashed) line corresponds to the prediction of equation \ref{disprel} ($\mu_{th} = 110$ Pa) with the interface pointing up (resp. down). The black dashed lines show the dispersion relation of shear waves: $\omega=k\sqrt{\mu/\rho}$.}
\end{figure}

Finally, we check the influence of gravity on the dispersion relation. We characterize a sample whose interface normal points upwards or downwards. In the first case, gravity acts as a restoring force on the free interface whereas in the second it tends to deform it and can even make it unstable \cite{mora2014, pandey2018}. Figures \ref{fig3}c and \ref{fig4}b present the dispersion relations obtained for a sample with $\mu=95$ Pa and $h=0.99\pm0.05$ cm when the interface points respectively up or down. For such a sample $|G| = 1.02$, so that we expect gravitational forces to matter but remain below the instability threshold.
The model accurately predicts the influence of gravity as shown by the overlay of the red solid line (respectively cyan dashed line) corresponding to the prediction of equation \ref{disprel} ($\mu_{th} = 110$ Pa) with the free surface pointing up (respectively down). This shows that by tuning $G$ below the value of the instability threshold, we can control the dispersion of the fundamental mode.

\section{Discussion}
In this article, we use state-of-the-art measurement techniques to probe the propagation of surface waves in agarose gels with great accuracy, revealing the importance of finite thickness that leads to the occurence of multiple modes at low frequency as well as the existence of an apparent dispersion.
We quantitatively predict the dispersion relation using an elastic model including capillary forces.
In particular, we capture the role of capillarity
even at wavenumbers lower than expected from scaling arguments
in finite thickness configurations through an intricate balance between in and out-of-plane interfacial displacements and in very thin samples.
We confirm the validity of our approach by including gravity in the model and successfully testing it against experimental data.
The influence of gravity opens new perspectives: $G$ can be tuned to create materials in which the phase and group velocity have opposite signs, a sought-after property allowing perfect lensing \cite{pendry2000}.
Furthermore, $G$ also depends on depth enabling to tune the medium properties down to sub-wavelength scales to create elastic metamaterials \cite{brule2014}.

\begin{acknowledgments}
We thank Sander Wildeman, Claire Prada and Jacco Snoeijer for insightful discussions. PC and AE thank David Qu\'er\'e for his support.
\end{acknowledgments}

\appendix

\section{Rheology measurements}
\label{apprheol}
\begin{figure}[t]
\centering
\includegraphics[width=1\linewidth]{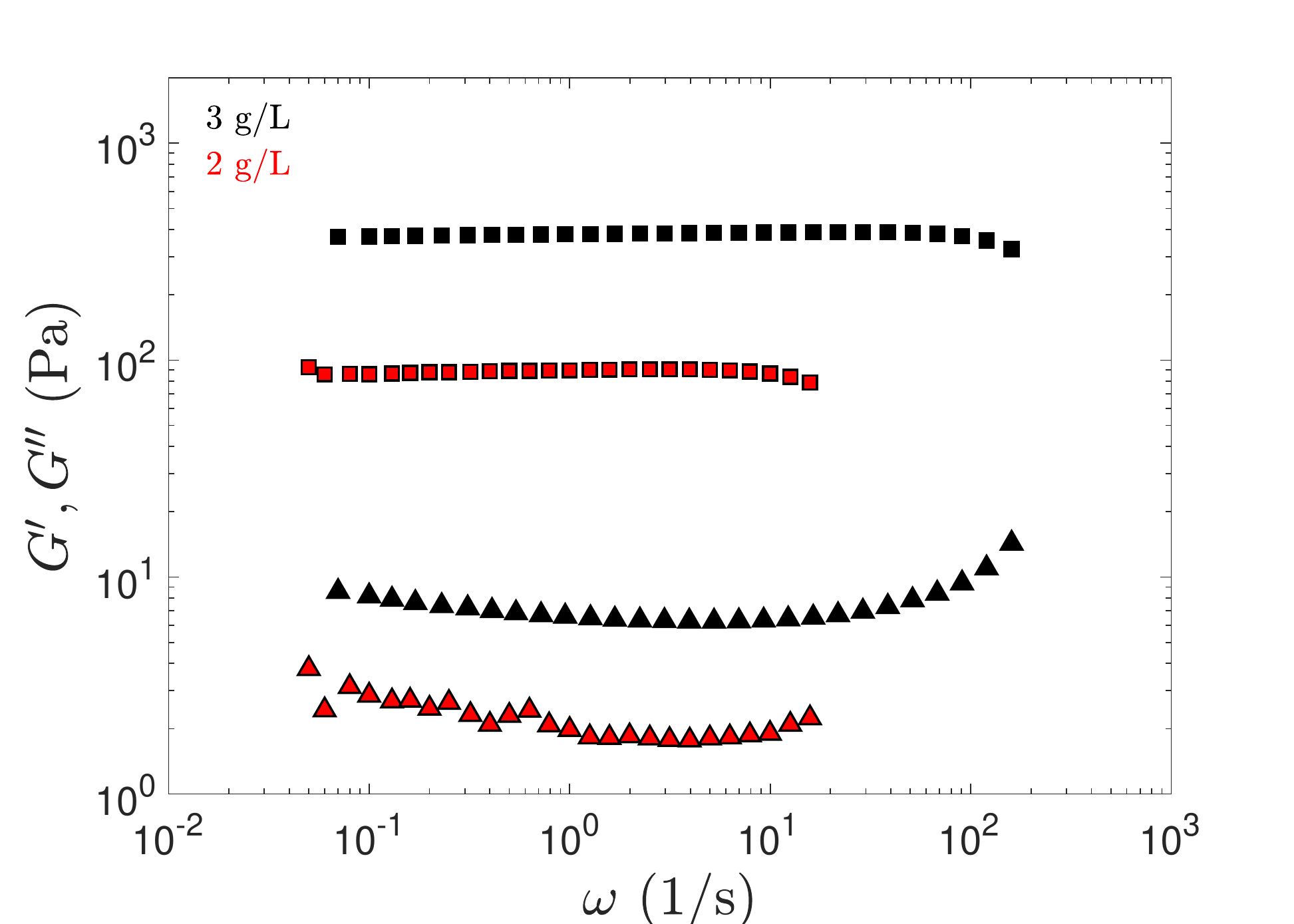}
\caption{\label{fig1SI} Storage ($G^\prime$, squares) and loss ($G^{\prime\prime}$, triangles) moduli of agar hydrogels with concentration of 2 and 3 g/L plotted as a function of the pulsation $\omega$.}
\end{figure}

We determine the rheology of the gels using a rheometer (Anton-Paar MCR501) in plate-plate configuration. We measure the shear modulus $\mu(\omega) = G^\prime + iG^{\prime\prime}$ for pulsations ranging form 0.05 to 100 rad/s at a fixed strain of 0.1\% and report the results in figure \ref{fig1SI}. In the probed range, both $G^\prime$ and $G^{\prime\prime}$ are constant and $G^\prime$ is typically one order of magnitude larger than $G^{\prime\prime}$.

\section{Role of evaporation}
\label{appevap}
\begin{figure*}[t]
\centering
\includegraphics[width=1\textwidth]{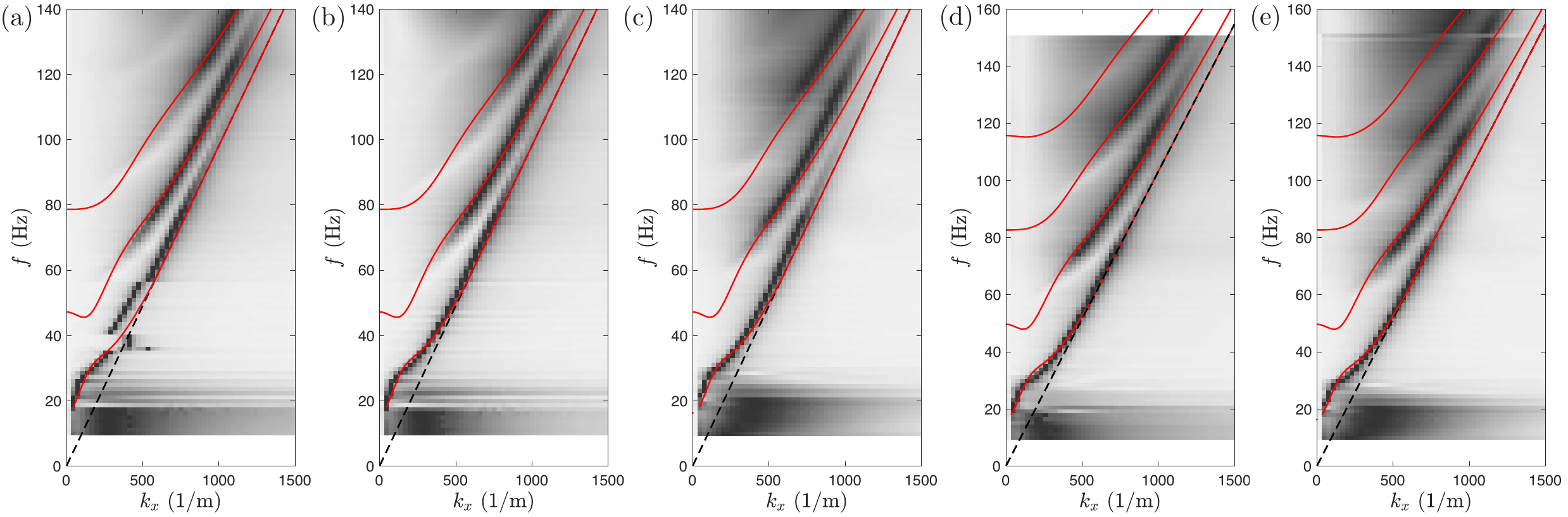}
\caption{\label{fig2_app} Dispersion relation measured from a gel with $\mu = 380$ Pa and $h = 0.98 \pm 0.05$ cm within 1h of reticulation (a), one (b), and three hours (c) after the first measurement. The red curves are obtained from equation 1 in the main text with $\mu_{th} = 380$ Pa. (d-e) Overlay of the dispersion maps for a gel with $\mu = 380$ Pa and $h=9.8$ mm and the dispersion curves predicted from equation 1 in the main text ($\mu_{th} = 420$ Pa) for actuators with width (d) 4.5,  and (e) 14 mm.
}
\end{figure*}

Drying and more generally ageing is a major concern in hydrogels.
We use our gels just after reticulation is complete and the gel has reached room temperature. A measurement typically takes less than one hour and gels are discarded after they are measured. We are thus confident that there is no macroscopic skin at the gel surface, but there might be a small gradient of properties due to the slow drying occurring after the gel preparation. As an independent check, we measured the dispersion relation of a gel sample: within one hour of reticulation (fig. \ref{fig2_app}a), one hour after the first measurement (fig. \ref{fig2_app}b), and three hours after the first measurement (fig. \ref{fig2_app}c). We observe no significant change between the dispersion relation of figures \ref{fig2_app}a-b (except for an input signal error between 35 and 55 Hz in figure \ref{fig2_app}a). Yet we notice, in figure \ref{fig2_app}c, an increase of the local slope of the branches, an observation compatible with an increase in gel shear modulus that can be attributed to evaporation. Although evaporation occurs in our system, it is not a limiting parameter in our experiments.

\section{Finite source size}
\label{appfinitesize}
The finite size of the actuator could have an influence on the results. Indeed, the measured wave fields depend on both the response of the material and the ability of the source to generate waves at a given frequency. We probed the effect of source size by using two PMMA strips of different width. Figure \ref{fig2_app}d-e show the dispersion relations for a gel with $\mu = 380$ Pa and $h = 9.8$ mm obtained by using actuators with width 4.5, and 14 mm. The two dispersion relations are almost identical showing that our results are independant of the size the actuator in the range of parameters that we consider.

\section{Effect of inclusions}
\label{appinclus}
The presence of micro-particles can modify the material properties of the gels. We extracted the particle concentration from DIC images by binarizing the image to find the area corresponding to bright pixels. Knowing the thickness of the sheet ($\simeq$ 200 $\mu$m) and the size of the particles ($\simeq$ 10 $\mu$m), we obtain the particle concentration $\chi = 0.14\%$.
We deduce the density of the gel with inclusions, $\rho_{eff} = (1-\chi)\rho + \chi\rho_p = 1000$ kg/m$^3$.
Then, we determine the effect of the inclusions on the gel shear modulus using Eshelby theory \cite{eshelby1957}. We assume the inclusions are rigid as the shear modulus of the particles is much larger than that of the gel. The effective shear modulus, $\mu_{eff}$, in the case of dilute spherical inclusions is given by: $\mu_{eff} = \mu(1 - B\chi)$ where $B = - 15(1-\nu)/(2(4-5\nu))$ with $\nu$ the Poisson ratio. For the results shown in figure 2 of the main text ($\mu = 95$ Pa, $\nu = 0.5$), we get $\mu_{eff} = 95.3$ Pa.
The presence of the inclusions has a negligible effect on the density and shear modulus of the gel. We don't expect the results to be modified and neglect their influence.

\section{Modeling: derivation of the dispersion relation and of the displacement field}
\label{appmodeling}
We consider plane waves propagating along the $x$ direction in an infinite 2D plate of thickness $h$, density $\rho$ which elastic properties are characterized by the Lam\'e coefficients $\lambda$ and $\mu$. We use Helmholtz theorem to separate the displacement field $\bm{u}$ in a longitudinal curl free contribution $\bm{u}_l$ and in a transverse divergence free contribution $\bm{u}_t$. The longitudinal part can be described by a scalar potential $\Phi$ and the transverse part by a vector potential $\bm{H}$.
\begin{equation*}
\bm{u} = \bm{u}_l + \bm{u}_t = \nabla \Phi + \nabla \times \bm{H}.
\end{equation*}
Both $\Phi$ and $H_y$ verify a wave equation \cite{landau1986}:
\begin{equation*}
\nabla^2\Phi -\frac{1}{c_l^2}\frac{\partial^2 \phi}{\partial t^2} = 0,\, \nabla^2 H_y -\frac{1}{c_t^2}\frac{\partial^2 H_y}{\partial t^2} =0
\end{equation*}
where $c_t = \sqrt{\frac{\mu}{\rho}}$ and $c_l=\sqrt{\frac{\lambda+2\mu}{\rho}}$ are, respectively, the shear and longitudinal wave speeds.
We look for solutions of the form $\Phi = f(z)e^{i(kx-\omega t)}$ and $H_y = ih(z)e^{i(kx-\omega t)}$ where we write $k$ in place of $k_x$.
By substitution in the wave equations, we obtain:
\begin{equation*}
\frac{\partial^2 f}{\partial z^2} + \alpha^2 f = 0, \, \frac{\partial^2 h}{\partial z^2} + \beta^2 h = 0
\end{equation*}
where $\alpha^2 = \frac{\omega^2}{c_l^2} - k^2$ and $\beta^2 = \frac{\omega^2}{c_t^2} - k^2$.
We deduce the form of the solution of $f$ and $h$ and write the expressions for $u_x$ and $u_z$:
\begin{multline*}
u_x = i[k(A\sin\alpha z +B\cos\alpha z)+ \\
\beta(C\cos\beta z-D\sin\beta z)]e^{i(kx-\omega t)}
\end{multline*}
\begin{multline*}
u_z = [\alpha(A\cos\alpha z -B\sin\alpha z)+ \\
k(C\sin\beta z +D\cos\beta z)]e^{i(kx-\omega t)}.
\end{multline*}
From the displacements, we obtain the stresses $\sigma_{xz}$ and $\sigma_{zz}$:
\begin{equation*}
\sigma_{xz} = \rho c_t^2 \left(\frac{\partial u_x}{\partial z}+ \frac{\partial u_z}{\partial x}\right)
\end{equation*}
\begin{equation*}
\sigma_{zz} = \rho c_l^2\frac{\partial u_z}{\partial z}+\rho(c_l^2-2c_t^2)\frac{\partial u_x}{\partial x}.
\end{equation*}
We have now determined all quantities to express the boundary conditions.
At the bottom of the sample, we assume that the gel is bounded to the container:
\begin{equation*}
u_x(z=-h)=u_z(z=-h)=0.
\end{equation*}
At the free surface, assuming small deformations to linearize the boundary conditions at $z=0$ and taking advantage of the incompressibility of the hydrogels ($c_l\rightarrow\infty$) that allows to compensate gravity by a pressure field, we impose:
\begin{equation*}
 \sigma_{xz}(z=0) = 0,\, \sigma_{zz}(z=0) = \gamma \frac{\partial^2 u_z}{\partial x^2}-\rho g u_z.
\end{equation*}
The four boundary conditions yield four equations, involving the constants $A$, $B$, $C$ and $D$, that can be recast in matrix form:
\begin{widetext}
\begin{equation*}
\begin{bmatrix}
-k\sin\alpha h & k\cos\alpha h & \beta\cos\beta h & \beta\sin\beta h \\
\alpha\cos\alpha h & \alpha\sin\alpha h & -k\sin\beta h & k\cos\beta h \\
2k\alpha & 0 &0 & k^2-\beta^2 \\
k^2 \alpha(\gamma+\rho g /k^2) & \rho c_t^2(k^2-\beta^2) & 2\rho c_t^2k\beta & k^3(\gamma+\rho g /k^2)
\end{bmatrix} \cdot\\
\begin{bmatrix}
A \\ B \\ C \\ D
\end{bmatrix}
=0.
\label{matrixeq}
\end{equation*}
\end{widetext}
Waves propagate when there are non-trivial solutions to the above system, the dispersion relation is obtained by taking the determinant of the matrix.
This relation can be written in dimensionless form by introducing the variables $\tilde{k}=kh$ and $\tilde{\omega} =\omega h /c_t$ allowing one to obtain equation \ref{disprel}.
For any couple ($\omega$, $k$) that verifies the dispersion relation, we obtain the values of three of the constants $A$, $B$, $C$ and $D$ allowing to determine the displacement field up to a constant. We give the values of $A$, $B$, $C$ as a function of $D$:
\begin{widetext}
\begin{eqnarray*}
C = -D\frac{k^2 - \beta^2 + \alpha\beta e^{i\beta h}(e^{-i\alpha h}-e^{i\alpha h})-k^2e^{i\beta h}(e^{-i\alpha h}+e^{i\alpha h})}{k^2-\beta^2-\alpha\beta e^{-i\beta h}(e^{-i\alpha h}-e^{i\alpha h})-k^2e^{-i\beta h}(e^{-i\alpha h}+e^{i\alpha h})}\\
A = -\frac{1}{ik(e^{-i\alpha h}+e^{i\alpha h})}\left(C\left(\frac{k^2-\beta^2}{2\alpha}e^{i\alpha h}-\beta e^{-i\beta h}\right)+D\left(\frac{k^2-\beta^2}{2\alpha}e^{i\alpha h}+\beta e^{i\beta h}\right)\right)\\
B = A +i\frac{(\beta^2-k^2)(C+D)}{2\alpha k}.
\end{eqnarray*}
\end{widetext}

\section{Lam\'e modes}
\label{appLame}
Lam\'e modes are a special solution of the Rayleigh-Lamb equation obtained when $\omega = \sqrt{2}kc_t$. They correspond to the propagation of pure bulk shear waves at a 45$^\circ$ angle to the plate axis and to the maximum of normal displacement at the surface of the plate \cite{chati2011}.
One can show by substituting the expression of Lam\'e modes in equation \ref{disprel} that it reduces to the Rayleigh-Lamb equation for symmetric modes of a plate with thickness $2h$ in the absence of capillarity and gravity. Thus, the maxima of the normal displacement at the free surface of the gels align on a line with slope $\sqrt{2}c_t$ when we do not take into account surface tension.




%
%
%
%


\begin{thebibliography}{30}
\bibitem{evap}
\BibitemOpen
Evaporation occurs over larger surfaces in the samples than during rheology measurements which leads to increased agarose concentrations and thus stiffer gels\BibitemShut {NoStop}
\end{thebibliography}
\end{document}